# Software Defined Adaptive MIMO Visible Light Communications after an Obstruction

**Peng Deng and Mohsen Kavehrad**
*Department of Electrical Engineering, The Pennsylvania State University, University Park, PA 16802, USA*
*pxd18@psu.edu*

**Abstract:** We experimentally demonstrate a software-defined 2x2 MIMO VLC system employing link adaptation of spatial multiplexing and diversity. The average error-free spectral efficiency of 12 b/s/Hz is achieved over 2 meters indoor transmission after an obstruction.

## 1. Introduction

Worldwide growth in wireless data traffic has led to the development of the next generation mobile communication networks to address several critical challenges, such as broadband capacity, spectral efficiency, energy efficiency, mobility coverage, and quality-of-services. Visible light communications (VLC) based on light-emitting diodes (LEDs) merges lighting and data communications in applications for their high energy efficiency, spectral efficiency, security and reliability [1]. VLC applications as shown in Fig.1 (a) are relevant to 5G networks, Internet-of-Things, WPANs, aerospace, automotive, healthcare industries, and ad-hoc networks. The phosphor-based white-light LED has a limited modulation bandwidth [2] that can be improved by using a blue filter and equalization of the driving circuitry [3]. Techniques such as high spectrally efficient modulation [4], wavelength division multiplexing (WDM) [5] and optical multi-input multi-output (MIMO) [6] have enhanced capacity of VLC system.

However, current MIMO VLC demonstrations feature the direct line of sight, fixed MIMO technique, offline signal processing and static ideal channel condition estimation. In practical indoor wireless mobile communications, any line of sight link may be blocked by an obstruction such as human shadowing, and the complex channel conditions are expected to be changed dramatically. Once any sub-channel condition becomes degraded due to line-of-sight blockage, MIMO VLC system using the fixed spatial multiplexing will suffer link interruption. MIMO VLC using spatial diversity can improve link reliability with tolerance to shadowing, but there's no gain in spectral efficiency even with good channel condition. Thus, we propose software defined adaptive MIMO VLC, that both modulation techniques and MIMO techniques are dynamically adapted to the changing channel conditions, for enhancing both link reliability and spectral efficiency. Software defined implantation can assist in enabling an adaptive and reconfigurable MIMO VLC system without hardware changes. In this paper, we experimentally demonstrate a real-time software-defined Single-Carrier M-QAM MIMO visible light communication system by using link adaptation of spatial multiplexing and spatial diversity. The average error-free spectral efficiency of adaptive 2x2 MIMO VLC achieved 12 b/s/Hz over 2 meters indoor dynamic transmission after an obstruction.

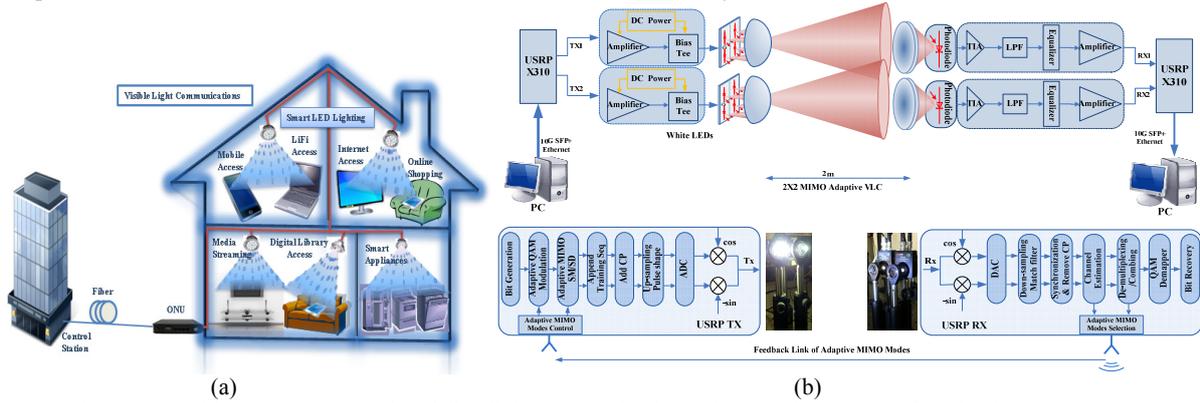

Fig. 1. (a) Architecture network with visible light communications. (b) Experiment setup for adaptive MIMO VLC.

## 2. Adaptive MIMO VLC experiment setup

Figure 1(b) presents a block diagram and experimental setup of this 2x2 M-QAM adaptive MIMO VLC system. The random binary data is generated in Labview and would be first split into two parallel streams, one for each transmitter (TX) channel. We develop adaptive MIMO modes control modules, which can adjust the optimal modulation formats and MIMO schemes to maximize the spectral efficiency and error performance according to the real-time MIMO channel conditions. There are two types of MIMO schemes for selection: Spatial Diversity (SD) MIMO and Spatial Multiplexing(SM) MIMO. The former one requires each Tx transmit the same data stream to improve the antenna array gain, while the latter one needs each LED send the different parallel streams to enhance the spectral efficiency gain. In each channel, the bit stream is mapped into M-ary quadrature amplitude modulation (M-QAM) and there are four types of modulation formats: 4-QAM,16-QAM,64-QAM and 256-QAM.Thus, there are eight adaptive MIMO modes available for link adaptation(SM-4,SM-16,SM-64,SM-256,SD-4,SD-16,SD-64,SD-256).Initilaly, we choose the default mode of SM-64(spatial multiplexing 64-QAM). Then channel training sequences are inserted into the symbol streams. After adding cyclic prefix (CP) and up-sampling, pulse shaping by a raised-cosine filter is employed.

The host computer transmits data streams to Universal Software Radio Peripheral (USRP) X310 via the 10G SFP Ethernet interfaces. After digital to analog conversion (DAC), the complex QAM streams are up-converted to RF frequency and generate the real analogue RF signals. At the analogue transmitter front-end, the output signal was first amplified and then superimposed onto the LED bias current by the aid of a bias-tee. The light source was a phosphorescent white LED that consists of four chips, each providing a luminous flux of 520 lm with a 120° full opening angle when driving at 700 mA. An aspheric lens was fixed to make sure the light transmits along the 2 meter straight direction. Light from the white LED was imaged onto a high-speed PIN photo-detector through an aspheric convex lens. The photocurrent signal was amplified by a low noise trans-impedance amplifier (TIA). A first-order analogue post-equalization was developed to extend the bandwidth based on phosphorescent white LED.

Subsequently, the received signals from each of the two receivers are routed to a RX channel of USRP X310. After down-converting to baseband and down sampling by a matched filter, the training sequences are acquired for synchronization and channel estimation. After removing CP, the received data streams are processed in a MIMO de-multiplexer. The final streams are then passed through the QAM demodulator to recover the original binary stream. Based on the channel estimation conditions, the adaptive MIMO modes selection module will calculate and select the corresponding optimum modes including modulation formats and MIMO schemes. We then update the three bit binary code for adaptive MIMO modes and feedback the adaptive mode code to the transmitter by using RF uplink. In Labview, we measured and analyzed the constellation diagram, EVM, BER and spectral efficiency performance for adaptive M-QAM MIMO VLC system.

## 3. Experimental results and discussion

In order to achieve adaptive MIMO visible light communication for indoor lighting environment, we first evaluate efficient estimation approaches to acquire the real-time channel conditions. Then we measured and compared the error performance and spectral efficiency of M-QAM MIMO VLC systems using spatial diversity and spatial multiplexing, respectively. Based on the adaptive MIMO modes selection criteria in terms of channel estimation, we demonstrated the adaptive real-time software defined MIMO VLC after a mobile obstruction.

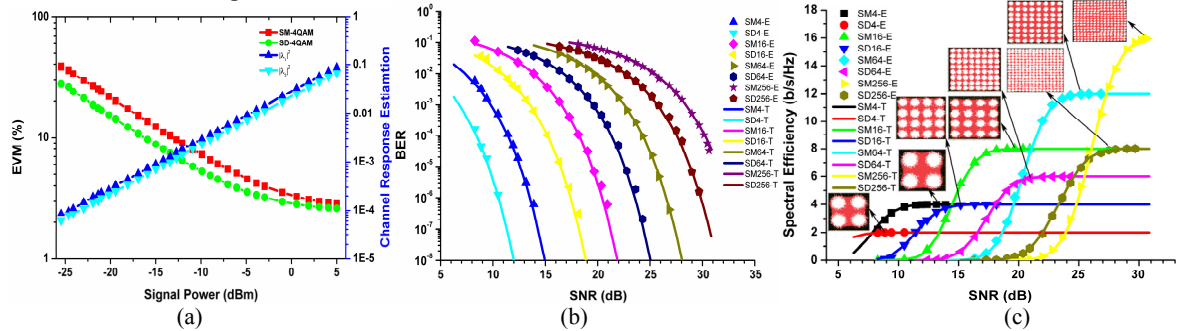

Fig. 2. (a) EVM and Channel Estimation against signal power for 2x2 MIMO VLC using spatial multiplexing. (b)BER and (c)spectral efficiency performance of M-QAM 2x2 MIMO VLC using spatial multiplexing and spatial diversity. Dots are experimental results and lines are theoretical results.

Figure 2(a) shows the measured EVM and channel response estimation against signal power for 2x2 MIMO 4-QAM VLC using spatial multiplexing and spatial diversity. The measured EVM can be used to estimate channel SNR, but it is dependent on the symbols number, constellation size and MIMO processing. Instead, we obtain channel SNR directly from channel response estimation matrix, where $\lambda_i$ is eigenvalue of channel estimation matrix H. We can see in Fig.2 (a) that each sub-channel eigenvalue of channel estimation has a linear relationship with signal power. Furthermore, original channel estimation matrix is similar for both spatial diversity and spatial multiplexing MIMO. Thus, we can acquire real-time sub-channel SNRs from eigen-values of channel estimation matrix regardless of MIMO schemes and modulation formats.

BER performance and spectral efficiency of 4-16-64-256 QAM 2x2 MIMO VLC using spatial multiplexing and spatial diversity are illustrated in Fig. 2(b) and Fig. 2(c), respectively. The dots are experimental measurements and lines are theoretical results. Insets are the corresponding constellation diagrams. The results show BER decrease significantly to $10^{-7}$ without nonlinear distortion by using single carrier $M$-QAM modulation. We can see that BER of spatial diversity system has about 3dB SNR gain compared to spatial multiplexing due to array gain. On the other hand, spatial multiplexing system has about 3dB gain in spectral efficiency compared with spatial diversity due to multiplexing gain. It is clear that the experimental measurements well match the theoretical results, which verify accurate channel SNR estimation from channel matrix in $M$-QAM MIMO VLC. Given the target BER threshold $BER_{tgt} = 10^{-3}$, we can establish the adaptive MIMO mode criteria in terms of channel SNR and singular value of channel matrix to maximize the spectral efficiency.

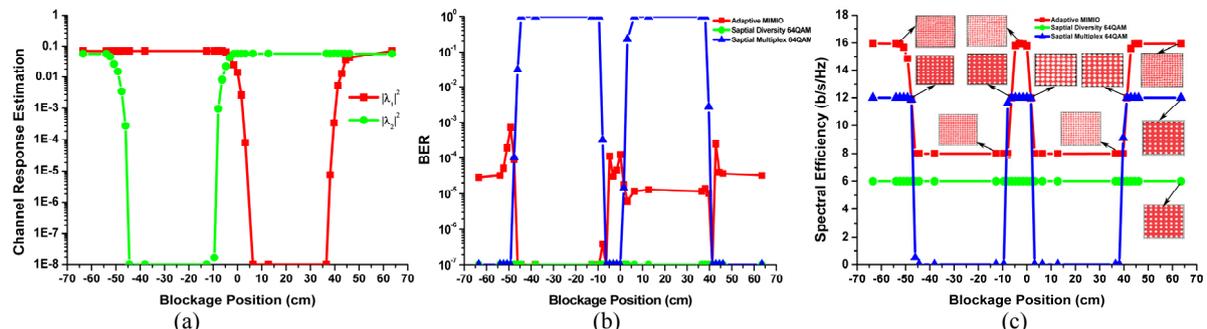

Fig. 3. Effect of link blockage on adaptive M-QAM 2X2 MIMO VLC using spatial multiplexing and spatial diversity. (a) Channel estimation against blockage position. (b) BER and (c) spectral efficiency against blockage position for adaptive M-QAM MIMO, spatial multiplexing 64-QAM and spatial diversity 64-QAM VLC.

Based on the established adaptive MIMO mode criteria in terms of channel estimation matrix, we demonstrate the real-time adaptive MIMO VLC system to maximize spectral efficiency under target BER threshold, after propagation beyond an obstruction. An obstacle with a diameter of 4.5 cm placed in the middle of transmission distance of 218 cm obstructs the line-of-sight light waves from two LEDs separated by 5 cm. We marked the central line between the two LEDs as a reference position at 0 cm. The obstacle was translated across the line-of-sight transmission direction from -65 cm to 65 cm, with a dynamic obstruction range of 135 cm. With the obstacle moving across LOS links, we measured the channel response of sub-channels and compared the error performance and spectral efficiency of adaptive M-QAM MIMO VLC with traditional MIMO VLC as shown in Fig. 3. The fixed 64-QAM MIMO VLC using spatial diversity and spatial multiplexing are inserted for comparison. Insets are the corresponding constellations of adaptive MIMO modes. The initial mode is SM-64(spatial multiplex 64-QAM). As we can see in the figures, as the obstacle moved across LOS from -65 m to 65 cm, the adaptive MIMO modes change to SM-256, SM-64 and SD-256 adaptively with BER less than $10^{-3}$. Results show that adaptive MIMO VLC can overcome the loss by blockage, thus recover the clear constellation diagrams after propagation beyond the obstruction. Adaptive MIMO VLC enhances the average error-free spectral efficiency to 12 b/s/Hz over 2 meters indoor transmission with an obstruction, compared with 7.7 b/s/Hz for SM-64 and 6 b/s/Hz for SD-64. The adaptive MIMO VLC will enhance spectral efficiency and error performance in a real-time combined lighting and communication environment such as transmission power distribution, environmental blockage and shadowing.

**Acknowledgments**
This work is support by the National Science Foundation (NSF) under Award #1201636 and Award #1160924.